\documentclass{edp-conf}
\usepackage{graphicx}

\DeclareRobustCommand{\element}{\relax\ifmmode\@tempswafalse
\else\@tempswatrue\fi\clearelargs\def\?{\phantom{0}}\@lement}
\def\@lement#1{\if#1[\expandafter\f@@dargs\else\druck@lement{#1}\fi}

\DeclareMathAlphabet{\mathsc}{OT1}{cmr}{m}{sc}
\def\testbx{bx}%
\DeclareRobustCommand{\ion}[2]{%
\relax\ifmmode
\ifx\testbx\f@series
{\mathbf{#1\,\mathsc{#2}}}\else
{\mathrm{#1\,\mathsc{#2}}}\fi
\else\textup{#1\,{\mdseries\textsc{#2}}}%
\fi}

\begin{document}

\TitreGlobal{AGN in their Cosmic Environment}

\title{Accretion and emission processes in AGN : \\ the UV-X 
connection} 
\author{ S. Collin}\address{DAEC, Observatoire de Paris-Meudon F-92190 Meudon ;
\email{suzy.collin@obspm.fr}}
\author{ A. Abrassart }\sameaddress{1}
\author{B. Czerny}\address{Copernicus Astronomical Center, bartycka 18, 
00-716 warsaw, Poland}
\author{ A.-M. Dumont }\sameaddress{1}
\author{ M. Mouchet }\sameaddress{1}
\secondaddress{University Denis-Diderot, F-75005 Paris}
\runningtitle{the UV-X-ray connection}
\maketitle
\begin{abstract}  The main characteristics of the average spectrum of radio quiet 
AGN in the UV and 
X-ray range are reviewed, and the 
emission mechanisms are discussed in the framework of accretion disk 
models, in particular the ``irradiated cold relativistic disk". It is 
shown that some problems arise in confronting the predictions of the model 
to the observations. 
We  propose an alternative model
in terms of a hot disk surrounded by a cold Compton thick medium. 
Finally we mention problems with remote 
regions of the accretion disk. 

\end{abstract}

\section{Introduction}

 It is now a paradigm to say that AGN and quasars are massive black holes fueled via 
an accretion disk. This model is mainly based on the dominance in their energy 
spectrum of a thermal feature, the ``Big Blue Bump" (BBB), 
extending from the optical to the EUV. However though much progress has been made in
 the last decade,
thanks to a bunch of new observations, the emission
mechanisms leading 
to the observed continuum and spectral features, as well as to their 
variability 
properties, are not easily related to the accretion disk. Moreover the 
physical processes taking place in the accretion disk are still not 
understood, neither at small nor at large distances from the black hole. 

 In the 
following we  first review the main characteristics in the UV and 
X-ray range (Sect. 2), then the 
emission mechanisms (Sect. 3). In Sect. 4 we recall how to integrate these 
 mechanisms in the framework of accretion disk models, and we
discuss some problems raised in confronting the predictions of the models 
to the observations. 
We  propose an alternative model to the ``relativistic disk" 
in Sect. 5. Finally in Sect. 6 we mention the problems of the remote 
regions of the accretion disk. Note that
some of the issues 
reviewed here have been discussed in more details by Collin-Souffrin \& 
Dumont (1997). 

\section{Observations in optical-UV-soft X}

We give here a very brief summary of some basic trends which are 
crucial for building a model of the central engine. X-ray observations    
are also reviewed in the same 
proceedings by Mouchet et al. For people interested 
in having more details, they can read Koratkar \& Blaes' (1999) excellent review.
We shall concentrate on radio quiet objects, which do not display the 
complications due to the presence of a non thermal component.

\subsection{The energy spectrum}

 Although there is a 
large range of properties, general trends can be deduced 
from the composite optical-UV spectrum of Francis et al. 
(1991) and Zheng et al. (1997), and from the soft X-ray spectrum of Laor et al. 
(1997). From 1eV to 10 eV  
the continuum is relatively ``flat", its spectral index 
$\alpha$ defined as 
$F_\nu\propto \nu^{-\alpha}$ being in the range -0.3 (for luminous quasars) 
to +0.5 ({\it note that 
$\alpha$ is generally positive for Seyfert galaxies and low luminosity quasars}. 
The  soft X-ray continuum below 1 keV seems to be simply the extension of the UV with 
a spectral index 
close to 1.5 (in $F_{\nu}$). It is this part of the spectrum, from 0.1eV 
to 1 keV,  which constitutes the Big Blue Bump. In Seyfert 1 nuclei the 
BBB contributes about three times less to the total power than in the previously 
largely used Mathews \& Ferland's (1987) continuum,
and  its peak
is located at a smaller energy, 10 eV instead of 50 eV (note that Mathews and 
Ferland  did not have 
at their disposal the HST and ROSAT results). The BBB is more important in 
luminous quasars. A noticeable fact concerning the UV continuum is the 
almost complete {\it absence of Lyman
discontinuity} (a weak absorption edge is sometimes observed but it is most 
probably due to the Broad Line Region or
to intervening intergalactic clouds with a redshift close to the emission 
one). The Warm 
Absorber inprints a few features in the soft 
X-ray continuum  (cf. Mouchet et al., in the same proceedings).

 Above 1 keV the 
spectral index is 0.7 in average, so in a majority of objects the continuum 
below 1 keV displays
 an excess when compared to the extrapolation of the
continuum in the 1-10 keV range.  Above 1 keV the spectrum can 
be decomposed in an
underlying power law with a spectral index $\sim 0.9$ and, superimposed 
on it, a broad emission line at 6.4 keV identified with
the K$\alpha$ line of weakly ionized iron and a ``hump'' peaking at about 
30 keV. The line shows an asymmetrical profile with an extended red wing 
(Nandra et al. 1997). In the two best studied objects, MCG-6-30-15 and 
NGC 3516, the line wing extends down to 4 keV.
The continuum has a cut-off with an e-folding energy of 100 to 200 keV.

A last important result is that the BBB luminosity is larger than the hard
X--ray luminosity, or at least of the same order.

\subsection{Variability}

The scaling of the size  $r$ in Schwarzschild 
radius $R_S$ with the variability timescale in day  
$t_{day}$ (assuming that the signal propagates at the speed of light) is:
\begin{equation}
r \sim 10^3\ t_{day}\ M_7^{-1}
\label{equ-time-size}
\end{equation}
 where $M_7$ is the black 
hole mass in 10$^7$ M$_\odot$. The same relation holds roughly for 
the distance between two regions and the time 
delay between their respective emissions (corresponding for instance to two 
spectral bands), provided the light 
curves are similar.

\subsubsection{Correlations between continuum interbands}

Variability studies are most important for understanding the mechanisms 
powering AGN. Unfortunately the conclusions one can draw from them are 
complex and sometimes controversial. A few of them are however 
firmly established concerning the timescales. First
the characteristic time variability of the UV flux (for variations of the 
order of one magnitude) scales with some power of the
luminosity: in Seyfert galaxies
it is typically of the order of a few days, while it is several months in high
luminosity quasars. Second the X-ray flux is more 
rapidly variable 
than the UV, with time scales of hours for Seyfert galaxies. Another well 
established result is that the optical-UV spectrum ``hardens when it brightens".

From the observations of a few objects intensively monitored with 
optical telescopes and/or IUE, HST, EUVE, ASCA, RXTE, BeppoSAX (MCG-6-30-15, 
NGC 3516, 4051, 4151, 5548, 7469), one can 
deduce more precise conclusions. 
For instance NGC 7469 shows 
variations of the UV and X-ray flux with similar large
amplitude and a time delay of the order of 2 days, the UV leading the 
X-rays in the flux peaks. But the X-ray flux
 displays in addition short term variations not seen in UV (Nandra et al. 
1998). For NGC 3516 the delay between UV and X is larger than 2 days
(Edelson et al. 2000). For NGC 4051 no optical variations were observed when 
strong X-ray variations were seen (Done et al. 1990).

 The UV flux variations drive the  optical flux
variations with a time delay $\le 0.2$d for NGC 
4151 (Edelson et al. 1996), $\le 0.15$d for NGC 3516 (Edelson et al. 
2000), and a delay of the order a day for NGC 7469
(Wanders et al. 1997, Collier et al. 1998).

Finally soft X-ray variations are 
generally larger than hard X-ray ones, but 
this is not always the case (Nandra et al. 1997). For NGC 3516, there were 
strongly correlated with no measurable lag ($\le$ 0.15d, Edelson et al. 2000).

 The fact that two light 
curves are very similar implies that there is a causal 
link between them. From these few results one can thus conclude that in 
Seyfert nuclei,

\noindent 1. the soft and hard X-rays are emitted by the same region.

\noindent 2.  the size of the 
UV emission region is larger than that of the X-ray, itself
 $\le 10^2M_7^{-1}$ $R_S$, 

\noindent 3.  the distance between the 
UV and the X-ray emission regions is  $\sim 10^3 M_7^{-1}$ $R_S$,

\noindent 4. the distance between the optical 
and the UV emission regions is $\sim 10^2 M_7^{-1}$ $R_S$.

For the two last results it is assumed that the causal link is functioning at 
the speed of 
light. If it would not be the case, the distance would have to be much smaller.
One deduces that {\it the optical 
and UV emission regions lie closer 
from each other than their dimension 
would let expect, or they simply are identical, and they both are at least one 
order of magnitude larger 
than the X-ray source}.

\subsubsection{Correlations between the continuum and the Iron K$\alpha$ line}

A few objects have been intensively monitored with ASCA and RXTE, revealing a 
complex and surprising behaviour of the line versus the underlying continuum. 
Iwasawa et al. (1996) have observed rapid variations of the Iron line 
profile and intensity in 1 to 10 ks in MCG-6-30-15 on the basis of an ASCA 
study. But this result has been recently questioned 
by Lee et al. (1999) and Chiang et al. (2000) who have studied with RXTE 
the spectral 
variability of MCG-6-30-15 and of NGC 5548, and have shown that the Iron 
line flux is found constant over time 
scales of 50 to 500 ks, while the underlying continuum displays large 
flux and spectral variations. Reynolds (1999) extended the work of Lee et 
al. on MCG-6-30-15, excluding a time delay between the line and the continuum 
in the range 0.5 to 50 ks, and suggesting that the line flux remains constant on 
timescales between 0.5 and 500 ks.

\section{Emission mechanisms}

\subsection{The BBB}

The shape of the BBB implies a thermal mechanism (i.e. where the bulk of the 
gas and not only a 
small fraction participates to the emission) with emission
regions spanning a temperature from a few 10$^4$K (optical emission) 
to 10$^6$K (EUV emission). This can be achieved either by an optically 
thick medium radiating 
locally like a black body or a 
modified black body, or by a medium optically thin at frequencies smaller 
than the peak of the Planck curve (the emission is thus due to free-free and 
free-bound processes).
The size of the emission region should satisfy the "black body limit":
\begin{equation}
r_{BBB} \ge 50 \left({\Omega \over 4\pi}\right)^{-1/2}\ T_5^{-2}\ 
M_7^{-1/2} 
\left({L_{\rm
UV}\over L_{\rm Edd}}\right)^{1/2}
\label{equ-rBBB}
\end{equation}
where $L_{\rm UV}$ is the luminosity of the BBB,  $L_{\rm Edd}$ the 
Eddington luminosity,
$\Omega$ the solid angle covered by the
emission region  (in the case of a disk it is 2$\pi$),
$T_5$ the effective temperature in units of 10$^5$K, and $r_{BBB}$ is
 in $R_S$.

Thus $r_{BBB}$ is marginally consistent with the dimension of the UV emission 
region deduced from variablity studies.

\subsection{Hard X-rays}

The absence of the electron-positron annihilation line and the 
existence of
the cut-off at about 100 keV argue strongly in favor of a thermal 
emission mechanism, with $ kT_{\rm e}$/$mc^2$ 
of the 
order of
0.1. The power law is generally believed to be due to Compton upscattering 
of UV photons (the BBB) by this hot medium. The hump and the 
Fe K line are attributed to reflection onto a 
Compton thick ``cold" medium (by ``cold" one means ``not highly ionized", 
which, in the case of radiative ionization, corresponds to a temperature 
$< 10^6$K). This medium is commonly identified with the BBB emission 
region.

\section{Emission and accretion processes close to the black hole}

One should therefore seek for models accounting for the presence of a 
very hot medium with a small size emitting the X-ray continuum, 
and of a ``cold'' medium with a larger size
emitting the BBB. There are also compelling evidences for a radiative 
interaction through inverse and direct Compton scatterings between these media.

The presence of an accretion disk is attested by several facts: 
first the black hole must be fueled, and second there are
evidences of a privileged direction (large gaseous disks, cones of 
ionized gas, collimated jets, Unified Scheme). So most naturally one tries to
account for the emission of the BBB and of the X-ray 
spectrum within the framework of accretion with angular momentum. If the 
flow is rotationally supported it will form an accretion disk.

The main problem with accretion disks remains the outward transport of angular 
momentum required to transport the mass inward. It is generally assumed to 
be  accomplished by a ``turbulent 
viscosity"  according to the $\alpha$ prescription proposed by Shakura and Sunyaev in 
1973. This prescription amounts to assuming that the viscous stress is equal to 
the total (gas + radiation) pressure multiplied by a parameter $\alpha$ of 
the order of unity, or equivalently that:
\begin{equation}
V_{rad}= \alpha c_s {H\over R}
\label{equ-alfa}
\end{equation}
where $V_{rad}$, $c_s$, and $H$ are respectively the radial velocity, the 
sound speed, and the scale height of the disk.
 For a detailed  
discussion of $\alpha$-disks one can refer for instance the 
book of Frank, King, and Raine (1992). Other prescriptions are under 
debate (cf. the paper by Hur\'e and 
Richard in these proceedings). 

\subsection{Generalities about accretion disks}

 There are several possible regimes according to the accretion rate 
 $\dot{m}$ expressed in terms of the Eddington rate, 
 $\dot{M}_{Edd}=L_{Edd}/ \eta
 c^2$,  where $\eta$ is the mass-energy conversion efficiency, generally assumed of 
 the order of 0.1. Let us summarize briefly their 
 characteristics. 

\medskip

\noindent 1. for $\dot{m} \sim  1$

Owing to the large accretion rate, the flow is optically thick. It 
is dense, so it radiates efficiently and stays cold. It is therefore 
radiating locally as a black body in the UV  and soft X-ray range. The 
inner regions of these 
disks are {\it supported by radiation pressure}.

If  $\dot{m} \ge  1$ the disks are inflated  and {\it geometrically thick}, 
their efficiency decreases with $\dot{m}$ (cf. Paczynski and Wiita 1980, 
Abramowicz et al. 1980, and further works). The emission spectrum depends strongly on 
the viewing angle, the spectrum being harder in the direction of the 
pole (cf. Madau 1988). According to Eq. \ref{equ-alfa} the radial 
velocity is large, and the {\it heat advected to the black hole is
larger than the heat radiated away}. As a consequence 
the mass-energy conversion efficiency $\eta$ can be much smaller than the 
canonical value of 0.1, and the disk has a relatively small luminosity for very
high  accretion rate. It is one of the 
reasons why these disks are not considered anymore. However they could be 
of interest for objects radiating near their Eddington luminosity, as it is 
proposed for Narrow Line Seyfert 1 for instance.

If $\dot{m} <  1$ the disks are geometrically thin (i.e. the scale height 
$H$ is much smaller than the radius $R$), and since the radial dynamics 
is  dominated by the gravitation of 
the black hole, they are in Keplerian motion. A majority of 
works have concentrated on these thin $\alpha$-disks, referred to as
{\it ``standard disks"}.
In this case {\it the amount of heat advected to the black hole is negligible}: 
\begin{equation}
Q_{adv}=  Q_{visc}-Q_{cool} \ll Q_{cool}
\label{equ-Qadv}
\end{equation}
where $Q_{visc}$ is the heat provided by 
gravitational release and $Q_{cool}$ is the heat radiated away.

\medskip

\noindent 2. for $\dot{m} <  \alpha^2$, i.e. typically $< 0.1$:

Owing to the small accretion rate, another solution can exist, where 
the flow has a small density and is 
not an efficient radiator. Moreover, if Coulomb collisions are the only 
process to transfer energy from the protons (to which the viscous energy is
given) to the electrons which radiate it away, the protons can stay very 
hot, not far from the Virial temperature, while the electrons will be
cooled to a lower temperature, leading to a two-temperature gas. However if
efficient plasma processes couple 
protons to electrons, they will keep the same temperature. This 
solution corresponds to disks {\it supported by gas pressure, hot, 
geometrically thick, and optically thin}. 
They can differ by the cooling processes, for instance whether or 
not soft photons are 
present to induce inverse Compton cooling (Ichimaru 1977,  Narayan \& Li 
1995, Abramowicz et al. 1995, Narayan et al. 1998). These disks 
are referred to as {\it ``Advection 
Dominated Accretion Flows"} or {\it ADAF}. When 
$\dot{m} <  \alpha^2$ both solutions (i.e. the standard disk and the ADAF) exist, 
without an obvious prediction how to switch between them.

\begin{figure}[h] 
\includegraphics[width=8cm,angle=0]{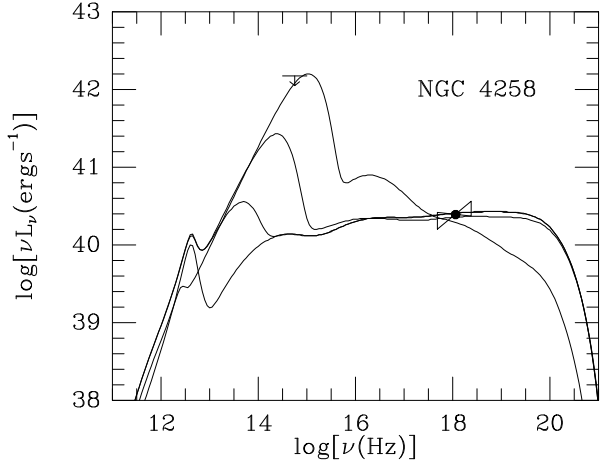}
\caption{Spectrum of an ADAF with a cold disk. The 4 curves correspond to 
different positions of the transition between the ADAF and the cold 
disk : infinity (no cold disk, pure flat ADAF 
spectrum), 500 $R_S$, 50$R_S$, 5$R_S$. The importance of the blue bump and 
its peak frequency increase when the transition radius decreases. Courtesy 
J.-P. Lasota}
\label{fig-ADAF}
\end{figure}

Much attention has 
been devoted these last years 
to ADAF. They are generally thought to be present in low 
luminosity objects, such as LINERs, which show a non resolved X-ray source 
but no optical-UV continuum. 
Indeed this is the kind of spectrum one expects from an ADAF. The ADAF can 
be surrounded by a thin cold disk located further from the black hole, 
where the cooling time becomes smaller than the viscous time (cf. Fig. 
\ref{fig-ADAF}). In this case 
the hot medium constituting the ADAF can be cooled by inverse Compton 
scattering of the soft photons produced by the cold disk. If a 
nuclear radio source is present, the synchrotron photons can play the same 
role in cooling the hot gas.

It is often admitted now that the Galactic 
Center (Sgr A*), the nucleus of NGC 4258, the 
nucleus of M87  and other nuclei of elliptical 
galaxies, are good 
examples of ADAFs (cf. Mezger et al. 1996, Lasota et al. 1996, Reynolds et 
al. 1996, 
for instance), though there are still strong debates on the subject. Since we 
are mostly interested by Seyfert galaxies 
and quasars in this review, we shall not spend more time on this vast 
problem, and we shall concentrate on the geometrically thin 
$\alpha$-disks, which correspond to $\dot{m}\sim 0.1$, as deduced for Seyfert 
galaxies and moderate quasars.

\subsection{Standard $\alpha$-disks}

  Most of the time it is assumed that
the disk is steady (at least locally), although this 
assumption is not always justified.
Since the radial coupling is much weaker than 
in the perpendicular direction, the computation of the structure of a 
geometrical thin disk
is more simple than that of a thick disk, as it is reduced to one dimension and 
a half (i.e. the radial structure can be decoupled from the vertical 
one). It has been shown that the inner part of 
$\alpha$-disks in AGN ($< 10^3R_S$) is dominated by radiation pressure and 
is {\it viscously and thermally unstable}. 

It is possible to get roughly the emission 
spectrum of a geometrical thin disk without computing the 
vertical structure and knowing the value of the viscosity parameter. The 
basic assumption 
underlying these models being that {\it the
gravitational energy is  entirely radiated locally}, to the first order the 
disk emits at each radius a blackbody spectrum at the
effective temperature determined by the local dissipation rate:
\begin{equation}
\sigma T_{eff}^4 =\ {3GM\dot{M}  \over 8\pi R^3}\ 
f(R)\ \ \  {\rm erg\ cm^{-2}\ s}^{-1}\
\label{eq-Teff}\
\end{equation}
where $M$ and $\dot{M}$ are the mass of the black hole and the accretion 
rate, and $f(r)$ is a correction factor depending on the angular momentum 
of the black hole. The total spectrum is given by integration on 
the disk surface : it is the well known spectral distribution 
$F(\nu) \propto \nu^{1/3}$ with a maximum at a frequency corresponding to the maximum 
effective temperature. {\it Thus the standard disk model predicts a much flatter 
slope than observed in the optical-UV range}, cf. Sect. 2.1. This has been 
also stressed in Koratkar \& Blaes' review (1999).

In reality things are more complex, because the disk does not radiate 
exactly like a black body, and many improvements should be added to 
 this rough calculation to get detailed spectral features. Unfortunately 
 these sophistications require to introduce some unknown physics into the 
 model (the way 
 viscous deposition varies with the height in the disk, for 
 instance). In present day computations,
 the disk is divided 
into a number of rings, whose vertical
structure and local emission are computed independently with a 
 self-consistent treatment of radiative transfer. This was performed for the
 first time for AGN by Laor \& Netzer (1989), assuming LTE. Ross, Fabian \& 
Mineshige (1992) relaxed the LTE assumption and took into account  
Comptonization but kept a constant vertical density. The most sophisticated models 
are those of Hubeny \& Hubeny (1998) who solve consistently the transfer 
and the vertical structure with a 
complete  non LTE treatment, introducing all 
relativistic corrections, but without 
taking into account Compton diffusions. 

One important result of these computations is that {\it the models are not able to 
account for the BBB emission up to 1 keV}, even in the case of maximally 
rotating black holes; in other words they predict a steeper slope 
than observed in the 0.2-2 keV  range.
Ross et al. (1992), found  that a non negligible fraction of 
the disk luminosity is radiated in the soft X--ray range, as a consequence of 
Comptonization, but they considered low masses and large accretion 
rates, which correspond to high temperatures of the disk (cf. 
Eq. 4.3), and are probably valid only for Narrow Line Seyfert 1
Galaxies but not for the 
majority of Seyfert nuclei and quasars. 

\subsection{Irradiated $\alpha$-disks}

According to the local blackbody emission,  $T$ is proportional to $R^{-3/4}$,  
so the UV emission is produced closer to the black hole 
than the optical emission. More precisely the mean radius $<R_{\lambda}>$ where 
the flux at a wavelength $\lambda$ is produced can be estimated from 
Eq. \ref{eq-Teff}:
\begin{equation}
<R_{\lambda}> \sim 1.2\times 10^{15} \dot{m}^{1/3} M_7^{2/3}\left({\lambda\over 
5000\AA}\right)^{4/3}{\rm cm} \sim 400 \dot{m}^{1/3} M_7^{-1/3}\left({\lambda\over 
5000\AA}\right)^{4/3}R_S
\label{eq-Teff}\
\end{equation}
where the numerical factor takes into account the averaging. 
Comparing this result with the observed time delays between UV and optical 
light curves we see that it is compatible with a causal link between the 
two emission regions propagating 
at the speed of light, but incompatible with the propagation of 
a perturbation in a viscous time, as would be expected with the standard 
model (the speed of propagation would then be equal to 
the $c_{sound} H/R$, which is many orders of magnitudes smaller than the speed 
of light). This led Courvoisier \& Clavel (1991) to question the standard model, 
and Collin-Souffrin (1991) to propose that the disk was actually 
irradiated by the X-ray continuum, and emitting as a result of both viscous and 
external radiative heatings. The discovery of the Iron line and 
of the X-ray hump by Pounds et al. (1990), implying reprocessing by a cold 
medium, is in agreement with this model of irradiated disk.

Then began the era of ``irradiated disks", consisting in an
X-ray point source located at a given height $H_X$ above the center of the disk 
(this model is sometimes referred to as the ``lamp-post model", cf. Fig. 
\ref{fig-modeles})
The optical-UV emission is only slightly modified with respect to the non 
irradiated disk: the new blackbody temperature is roughly equal to
$[(F_{visc}+F_{irr})/\sigma]^{1/4}$. Since the external flux $F_{irr}$
is proportional to $1/(R^2+H_X^2)^{3/2}$, the temperature dependence with the 
radius, $T\propto R^{-3/4}$, is not changed with respect to the viscous disk at large distances 
from  the black hole, $R \gg H_X$, and the spectral distribution stays 
$F_{\nu}\propto \nu^{1/3}$. To get a steeper spectrum, $F_{\nu}\propto 
\nu^{-\alpha}$ with $\alpha\ge 0$ as observed, {\it the height of the X-ray 
source should be comparable to the distance of the 
UV-optical emission 
region, which is much larger than the size of the X-ray source}. Thus rapid 
X-ray fluctuations will be erased by the light travel 
time towards the disk, as observed. But {\it the UV optical light curves should 
always lag behind the X-ray one, while the opposite is sometimes observed}, 
for instance in 
the case of NGC 7469. A prediction of the model is that the UV flux would 
be more variable than the optical, corresponding to a hardening of the 
spectrum correlated with a brightening, since for $R>H$,  $<R_{\lambda}> $ 
is still given by Eq. \ref{eq-Teff}, as observed. One should note however that in this 
model, to induce similar 
variation amplitudes in UV and X-rays, {\it  $L_X$ should be of the order of 
$L_{UV}$}, contrary to the majority of 
objects.

\subsection{The disk-corona model}

\begin{figure}[h] 
\includegraphics[width=16cm,angle=0]{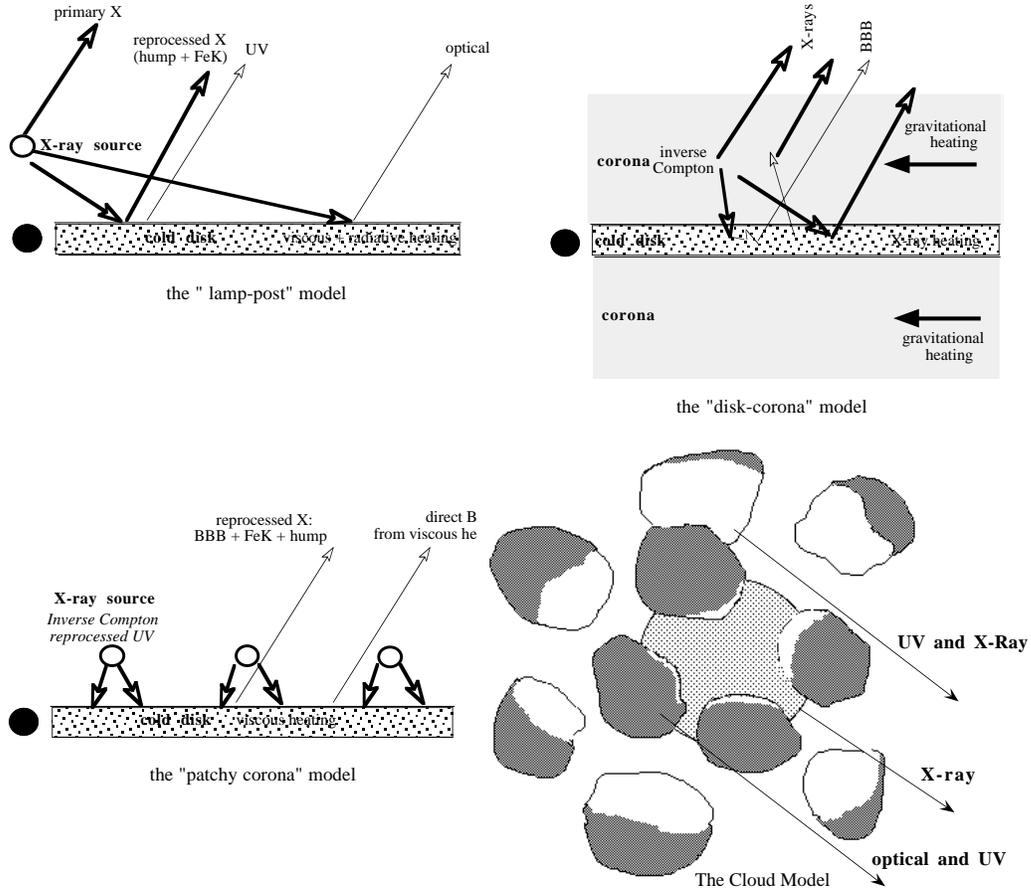}
\caption{Models for the UV-X emission.}
\label{fig-modeles}
\end{figure}

In the previous model the X-ray source is given  a priori without any 
physical background. A more physical model was proposed Haardt \& 
Maraschi (1991 and 1993),  consisting in
an optically thick cold disk sandwiched within an optically thin hot corona 
where a large
fraction of the dissipation is assumed to take place. The corona emits 
X-rays by inverse
Compton diffusions of  the soft UV photons coming from the underlying disk. 
The disk is
heated by the X-ray photons from the corona, and reprocesses this radiation as 
thermal UV (cf. Fig. \ref{fig-modeles}).  An advantage of this model is that it is radiatively 
self regulated and leads to a quasi universal spectral distribution. However it 
probably gives a much too strong Lyman discontinuity (Sincell \& Krolik 1997), though the 
physics of the transition region between the disk and the corona producing
the Lyman continuum is 
complex and still not well understood (Rozanska et al. 1999). 

The
ratio between the inverse Compton luminosity and the soft luminosity being 
of the order of unity, the model predicts a BBB/X-ray 
luminosity ratio smaller than observed. This is why Haardt, Maraschi \&
Ghisellini (1994) have proposed a variant, the
``patchy corona" (cf. Fig. \ref{fig-modeles}). The corona is not homogeneous, but is made of a few blobs, 
which could be due to the formation of magnetic 
loops storing energy and releasing it rapidly through reconnection like in 
solar flares.
Below a flare, the disk is heated to a higher temperature.  The rest of the 
disk radiates through viscous processes. One interest of 
the model is that the 
emission spectrum is independent of the fraction of gravitational power 
dissipated in the blobs, i.e. on the UV/X luminosity ratio, as it is observed.

Though giving a physical ground to the ``lamp-post model" and being able to 
account for very short X-ray variation timescales, the patchy corona 
suffers from the same drawbacks as the lamp-post. Indeed the X-ray 
flares
are located at only a few  
scale heights above the disk (Haardt et al. 1994), i.e. at a few 
Schwarzschild radii. This is the expected size of the variable UV emission region, 
assuming it is due to X-ray reprocessing; therefore it is much smaller 
than that
given by the observations.

\subsection{The Iron line and the irradiated disk model}

So far we have discussed some problems of the irradiated disk model linked 
with   the variations of the UV and X-ray fluxes. Let us now consider the 
Iron K line.
 
Since the discovery of the extended red wing of the Fe K line and of its short 
time variations, it has been proposed that the line is emitted by 
reprocessing  of the X-ray continuum at the surface of the disk, 
very close to the black hole, 
so that the broad red wing could be accounted for by gravitational 
redshift: it is the well known 
{\it ``relativistic disk"} introduced for Cyg X-1 by Fabian et al. 
(1989), now widely accepted for AGN, and used to model the line 
in data reductions. The line is assumed to be emitted at a rest energy of 6.4 
keV (i.e. it is a ``cold" line), the emissivity being parametrized as a 
power law $R^{-a}$. In several 
objects, in particular MCG-6-30-15 and NGC 3516, the index $a$ is very 
large (up to 8 in one state) unless one assumes that the emission is produced 
down  to a 
radius of the order of one $R_S$, which favors a Kerr geometry on 
a Schwarzschild one (Nandra et al. 1998). 

Obviously the emission of the Iron line and of the UV continuum are 
correlated with each other, as they are both due to the reprocessing of the 
X-ray continuum in the atmosphere of the irradiated disk. Therefore {\it the regions 
contributing the most to the variable fraction of the UV flux should give rise 
to an Iron line}. On the other hand the region contributing to the Iron 
line should produce an UV or an EUV excess of continuum. Actually Nayakshin 
et al. (2000) 
and  Nayakshin (2000) have shown that the emissivity of the line and the spectral 
distribution of the continuum reprocessed in the atmosphere of an 
irradiated disk depends  strongly on a ``gravitational parameter", which 
is a function of the ratio of the viscous to the X-ray heating of the disk. 
The  line emissivity and the UV spectral distribution
might therefore be
complex functions of $R$ and of the X-ray flux,  and one must wait for 
progress in the computation of the vertical structure of the disk, taking 
into account the irradiation, to get definitive answers to this question.

However we have seen that the optical-UV variable continuum is produced at a 
large distance from the black hole (at $R \sim$ 100 to 1000 
$R_S$), so whatever the details of the model, 
one  can expect that a fraction of the Iron 
line should be reprocessed in this region, leading to the emission of a 
narrow line at 6.4 keV, constant on time 
scales corresponding to the UV variation time scale. It is not clear 
whether a narrow core is always observed, but the broad line is, and we 
know that it stays constant 
during time scales of the order of days, while the underlying X-ray 
continuum varies.  So it is difficult to imagine that it is 
gravitationally broadened, 
and  the best way to account for the red wing is then
down Compton scattering in a Compton thick medium surrounding the 
source of line photons. We will propose 
this alternative model in the following section.

\section{The Comptonisation model}

Compton broadening of the Iron line has been suggested by 
Czerny et al. 
(1991). If the Iron line is intrinsically narrow because it is emitted by remote 
parts of the disk located at, say, $100
R_S$ from the black hole, it can be Compton broadened to the observed 
profile when crossing an ionized and relatively cold
Compton thick medium surrounding the disk: this is {\it Comptonisation by 
transmission} (cf. Misra \& Kembhavi 1998, Misra \& Sutaria 1999,
Abrassart \& Dumont 1999). But another way to get a broad Iron line is 
without a cold 
disk, but instead a hot disk surrounded by a cold Compton thick 
medium, so  the Iron line and the hump are emitted by the 
Comptonizing medium itself (cf. Collin-Souffrin et 
al. 1996, Abrassart \& 
Dumont 1998): it corresponds to {\it Comptonisation by reflection}. 
In Fig. \ref{fig-modeles} the ``cloud model" sketches such a 
possibility. In this figure a spherical symmetry has been assumed, but 
it is clear that the model can also correspond to an {\it axially symmetric thick disk 
geometry}.

This model is only phenomenological, and is not based on any physical ground. 
It is however possible to speculate on its nature. There are several 
observational proofs of the existence of an outflowing medium in AGN: the 
Broad Absorption lines, the fact that the UV lines are blueshifted with 
respect to the systemic velocity, the Warm Absorber which could be also in 
outward motion. There are also theoretical arguments for the 
existence of hydrodynamic winds or radiation driven flows close to the 
black hole. One can thus imagine that the Comptonizing medium 
is simply the basis of the Warm Absorber, for instance similar to the 
wind envisioned by 
Murray \& Chiang (1995). One interesting aspect of such a model would be 
that in this case the Iron line would be easily redshifted by Doppler 
effect from the 
position of the FeXXV or FeXXVI line (6.7 keV) to 6.4 keV or less, since the 
observer would be seeing mainly a line coming from the illuminated side of 
the clouds. 

 Since 
Mouchet et al. have discussed this issue in the same proceedings we
only give here a brief summary of the advantages and drawbacks of this 
model compared to the disk model. A more detailed discussion will be 
published elsewere (Abrassart et al. 2000). 

\subsection{The transmission model}

The transmission model has been rejected since the beginning by Fabian 
et al. (1995)  on 
the basis that to produce a weak (or no) Iron absorption 
edge (as it is observed), and at the same time to give the required amount of 
Compton broadening, the Comptonizing medium should be fully ionized, 
i.e. Iron must be stripped of all its electrons. To get such an 
ionization through all the Compton thick medium, the 
``ionization parameter" $\xi=L_{ion}/nR^2$ (where $L_{ion}$ is 
the  ionizing 
luminosity, $\sim L_{bol}$, and $n$ is the number density) should be $\ge 
10^6$.  Assuming 
that the Comptonizing medium is homogeneous (the most conservative case), 
one deduces for the distance between the Comptonizing medium and the X-ray 
source expressed in 
$R_S$, $r_{Compt}\sim 50 \dot{m} N_{25}^{-1}$, where $N_{25}$ is the 
column  density expressed in 10$^{25}$ cm$^{-2}$. Interestingly, 
this value does 
not depend on the mass of the black hole. We see that, provided the 
Eddington ratio is close to unity (and it is likely the case for 
MCG-6-30-15, whose mass is probably not much larger than 10$^6$M$_{\odot}$, cf. 
for instance Reynolds 1999), the distance of the Comptonizing region is 
marginally consistent with the assumption of a non relativistically 
broadened line which is broadened in the Comptonizing medium.

Reynolds \& Wilms (1999) discussed more quantitatively 
this model and concluded that 
there is no possibility left in the parameter space when one takes into 
account all the constraints set by the observations. 
However in Abrassart et al. (2000), it is shown that the model is viable, 
provided that the inner source is only partially covered by the 
Comptonizing medium ($\sim 90\%$). These computations are based on a 
photoionization-transfer code developed to treat Compton 
thick hot media (Dumont, Abrassart \& Collin, 2000). Moreover the
 big advantage of the model is to {\it erase any Lyman discontinuity} which 
would be produced inside the Comptonizing medium, without appealing to 
a fine tuning of the parameters.

\subsection{The reflection model}

The reflection model is different in that it does not require a very 
high degree of ionization, as the Iron line is produced in the 
Comptonizing 
medium itself. Typically $\xi$ should be of the order of 10$^3$, and actually a 
 range of ionization parameters is required to fit the data (Abrassart 
 1998).
The temperature of the medium is of a few 10$^5$K. {\it The 
Comptonizing medium is thus also emitting the BBB}. This is precisely the 
quasi spherical model proposed by Collin-Souffrin et al. (1996) and by 
Czerny \& Dumont (1998), to account 
for the UV-X spectrum of AGN. This model is made of a central X-ray source 
surrounded by a quasi spherical ensemble of Compton thick clouds with 
a large partial coverage 
factor. The observed X-ray continuum is a mixture of the primary 
continuum and of the continuum reprocessed by the illuminated side of the 
clouds, while the BBB is produced both by intrinsic emission of the 
clouds heated by the X-ray source and by reflection on the illuminated 
side of the clouds (cf. Fig. \ref{fig-modeles}). Like in the disk-corona model 
the X-ray source can be the result of Compton upscatterings of the UV photons 
emitted by the surrounding clouds. The radius of 
the Comptonizing-emitting medium 
is $r_{Compt}\sim 40 \dot{m} n_{14}^{-1/2}M_7^{-1/2}\xi_3^{-1/2}$ 
where $n$ is expressed in 10$^{14}$ cm$^{-3}$ and $\xi$ in 10$^3$. 

This model 
has been applied by Abrassart (1998) to MCG-6-30-15, but it predicts 
an excess of highly ionized Iron emission at 6.9 keV. To get rid of this 
component, it is necessary to assume that the medium is made of a mixture 
of a mildly ionized gas with a high density ($\xi\sim 10^2$) and a highly 
ionized gas with a low 
density 
(but in this case the gas has not to be fully ionized, 
since it can emit a small fraction of the Iron line, 
$\xi\sim 10^4-10^5$ is acceptable). The Iron line and the BBB are emitted 
by both media, and the Iron line is Compton downscattered in the highly 
ionized medium. 

\subsection{What are the advantages of the Comptonization model on the 
relativistic disk?}

A main difference with the disk model is that the region where the Iron line is 
emitted is located farther from the black hole. This 
is in agreement with the absence of short term variations of the line 
correlated  with variations of the continuum. Another advantage of the 
Compton model is to 
allow for the emission of the UV and of optical bands at similar 
distances, of the order of 100$R_S$, contrary to the disk model,
and in agreement with the relatively 
large time delay between the X-ray and the UV flux variations, and the 
small delay between the UV and the optical flux variations. It is also 
possible with this model to explain why in some cases the UV light curve 
lags behind the X-ray one, and it is the opposite in other cases. As 
discussed above, the UV and the X-ray spectra are due to the 
superposition of radiation from the central source (hard X-rays), radiation 
from the 
inner sides of the clouds further from the observer (soft X-rays and UV), 
and radiation from the outer side of the clouds closer to the 
observer (optical and UV). Moreover the photons are scattered several times in the 
Comptonizing medium, which complicates their travel. If the radiation emitted 
by the outer sides of the 
clouds dominates the UV (this depends on the coverage factor and on the 
thickness of the medium), one can well imagine that the UV photons will lead the 
soft  X-rays which are emitted further. On the other hand, this model 
allows for the possibility of clouds obscuring the line of sight in a 
random way. This process is studied by Abrassart \& Czerny (2000), 
and can also account for some variations. 
Finally a major difference with the disk model is that it gives naturally 
a $L_{UV}/L_X$ ratio larger than unity since the coverage factor of the 
``cold" medium is large. Finally it allows to get rid of the Lyman 
discontinuity, a long standing and not solved problem. 

\section{What happens to the disk far from the black hole?}

In an $\alpha$-disk the self-gravity overcomes the vertical 
black hole gravity beyond 
a few 10$^3 R_S$, as 
shown by Collin \& Hur\'e (1999). Consequently the disk becomes 
gravitationally unstable (Toomre 1964, Goldreich \& Lynden-Bell 
1965). The $\alpha$-prescription is obviously not valid if the 
disk is gravitationally unstable. Moreover the 
viscous time for transporting the gas towards the black hole is then 
larger than the lifetime of the active nucleus.

What is the structure of the disk in this region, and how is the outward 
transport of angular momentum ensured? We know that much farther from 
the  center, at about 10$^6\ R_G$, the disk is
globally gravitationally unstable, and the supply of gas can be achieved by 
gravitational torques or by global non axisymmetric 
gravitational instabilities (cf. Combes in these proceedings). But the
problem of the mass transport in the intermediate region where the 
disk is locally but not
globally self-gravitating is still not solved. 

Two solutions have been proposed. The first consists in
a disk made of  marginally unstable and
randomly moving clouds with large bulk velocities, where the tranfer of angular
 momentum is provided by cloud 
collisions (Begelman, Frank, \& Shlosman 1989). A more elaborate model 
is proposed for NGC 1068 by Kumar (1999) of a 
disk made of gas clumps undergoing gravitational interactions on one 
another. In this case the disk would be {\it geometrically thick}, and 
actually close to the picture of the torus invoked in the Unified Scheme. 
However the observations of the purely keplerian velocities of the maser 
spots lying at about 10$^5\ R_G$ from the black hole in NGC 4258 (Neufeld 
\& Maloney, 1995) 
argue strongly for a geometrically thin disk at this distance of the black 
hole, and are 
actually interpreted as setting on the top of a {\it warped thin disk}.
Collin \& Zahn (1999) have therefore adopted the opposite view that if unstable 
clumps
begin to collapse, the collapse will continue until protostars are formed,
which then accrete at a high 
rate and acquire masses of a few 
tens of M$_{\odot}$, leading to a ``starburst" of massive stars (not 
comparable in size with real starbursts). In this case the disk 
would stay geometrically thin, and its gravity would be
dominated by the gas. 
The mass transport would be ensured by supernovae or by outflows from 
stars, which at the same time 
would
contribute to the enrichment in heavy elements observed in the BLR and in 
the BAL QSOs (Hamann \& Ferland, 1992). A starburst at a larger scale 
could also be induced in the shocked gas of the 
supernova remnants. In this meeting
were indeed reported several evidences for the existence of  
starbursts in the very central region of Seyfert galaxies (cf. Terlevich's 
talk, for instance). The intermediate 
case  of a highly 
turbulent inhomogeneous thick
disk made of interacting clouds and containing massive young stars is also 
a
possible solution.

\section{Conclusion}

We have shown that besides raising several theoretical challenges, such 
as to assess the 
real state of the accretion disks close to the black hole (hot and thick, 
cold and thin?...) and to compute in a self-consistent way their structure 
and emission spectrum, the disk models are difficult to reconcile with 
the  observations, in particular with the timing 
properties. We have proposed an alternative model, where the disk is hot 
and thick, and is
surrounded by an inhomogeneous Compton thick medium, which reprocesses the 
X-ray continuum, emits the BBB and the broad Iron line, and
satisfies at the same time the requirements of time variations. We suggest that  this
medium could be in outward motion and linked with the Warm Absorber. More 
observations on the variations of the Iron line profiles and of the UV and X-ray
flux are needed to confirm or infirm these models. Finally we have 
also mentioned problems concerning the remote 
regions of the accretion disk.


\end{document}